\documentclass[12pt]{article}
\usepackage[utf8]{inputenc}
\usepackage[english]{babel}
\usepackage{amsmath}
\usepackage{amsfonts}
\usepackage{amssymb}
\usepackage[dvipsnames]{xcolor}

 \usepackage{bm}

\def\lb{\label}
\def\be{\begin{equation}}
\def\ee{\end{equation}}
\def\bea{\begin{eqnarray}}
\def\eea{\end{eqnarray}}

\begin{document}

\begin{titlepage}

\vspace*{0.7cm}

\begin{center}

{\LARGE\bf Generalization of the Bargmann-Wigner }

\vspace{0.3cm}

{\LARGE\bf approach to constructing
 relativistic fields}\footnote{Based on a talk presented by A.P. Isaev at the International Conference
on Particle Physics and Cosmology \\(professor V.A. Rubakov memorial conference), October 02-07, 2023, Yerevan, Armenia.}

\vspace{1.7cm}

{\large\bf I.L.\,Buchbinder$^{1}$\!\!, \,\,
S.A.\,Fedoruk$^1$\!\!, \,\,
A.P.\,Isaev$^{1,2}$\!\!\!, \, \, M.A.\,Podoinitsyn$^{1}$}

\vspace{1.7cm}

\ $^1${\it Bogoliubov Laboratory of Theoretical Physics, JINR,
141980 Dubna, Moscow region, Russia}, \\
{\tt buchbinder@theor.jinr.ru, fedoruk@theor.jinr.ru, isaevap@theor.jinr.ru, mpod@theor.jinr.ru}

\vskip 0.5cm

\ $^2${\it Physics Faculty,
Lomonosov Moscow State University, Russia}

\end{center}

\vspace{1.5cm}

\begin{abstract}
We review the method for constructing local relativistic fields corresponding to the Bargmann-Wigner wave functions that describe
the unitary irreducible representations of the $4D$ Poincar\'{e} group.
The method is based on the use of the generalized Wigner operator
connecting the wave functions of induced representations and local relativistic fields.
Applications of this operator for constructing
massive local relativistic fields as well as massless helicity local fields and massless local infinite spin fields are considered.
\end{abstract}

\vspace{3.5cm}


\vspace{0.5cm}


\vspace{1cm}

\end{titlepage}

\setcounter{footnote}{0}
\setcounter{equation}{0}

\newpage

\setcounter{equation}0
\section{Introduction}
\quad \ The Poincar\'{e} group is the mathematical expression of full relativistic symmetry.
Therefore, the study of various aspects of the Poincar\'{e} group and
its representations is
closely related to describing the properties of fundamental objects in modern physics. It is sufficient to say that just irreducible unitary representations
of the Poincar\'{e} group underly the classification of elementary particles in mass and spin.

The fundamentals of describing unitary irreducible representations of the $4D$ Poincar\'{e} group were laid down in the
eminent paper by Wigner \cite{Wig39}\footnote{Earlier, this subject, in a slightly less general form,
was also studied by Majorana \cite{EM} and Dirac \cite{P.A.M}.}. The approach to describing irreducible representations of the Poincar\'{e} group was further developed in the works by
Wigner and Bargmann \cite{Wig48,BarWig48}, where a general construction of such representations was proposed.

As is well known, irreducible unitary
representations of the Poincar\'{e} group are divided into two classes: massive and massless. Massless representations in turn
are subdivided into helicity representations and representations with continuous (infinite) spin.
It is generally believed that only helicity representations
are physically acceptable. However, recently continuous (infinite) spin representations
have attracted attention as well
(see e.g. \cite{BekSk,BuchIFKr,ShTr1,ShTr2}  and the references therein).

Following Wigner, all unitary irreducible representations of the Poincar\'{e} group can be constructed as induced ones from unitary irreducible representations of
the test momentum stability subgroup (Wigner little group). As it is known, these representations are realized on the Bargmann-Wigner wave functions \cite{Wig48},
\cite{BarWig48}. However, in coordinate space, these functions are non-locally transformed under the action of the Lorentz group.
Therefore, in their initial form,
the Bargmann-Wigner wave functions can not be used to construct relativistic field theories which are formulated
in terms of local fields. The problem of constructing
bosonic relativistic fields on the basis of the Bargmann-Wigner wave function was discussed in our recent papers \cite{BIPF,BIFPP,IsPod0},
where the method of deriving the corresponding relativistic fields was developed.
We were used the ideas of papers \cite{StW,Weinb}.
The central object of such a construction is the generalized Wigner operator that
transforms the Bargmann-Wigner wave functions into local relativistic fields. The form of this operator is different for different classes of representations:
it is a matrix acting on tensors for massive representations,
{an integral
operator} for representations with continuous (infinite) spin, and for helicity ones it is given in terms of
distributions.  In fact, this generalized Wigner operators provide the most general {relativistic equations and}
expressions for local relativistic  fields corresponding to unitary irreducible
representations of the Poincar\'{e} group. The paper under consideration is a brief review of our approach.

The work is organized as follows. In Section \ref{Se2}, we recall the basic notion
concerning the irreducible representations of the Poincar\'{e} group.
Subsection \ref{Se31} is devoted to describing the general construction of the induced Wigner representations.
In Subsection \ref{masrep}, we discuss in detail the case
of massive representations, their realization in terms of local relativistic fields and
free equations of motion for these fields. Subsection \ref{Se33}
is devoted to the explicit construction of Bargmann-Wigner wave functions
for massless representations with continuous (infinite) spin. In Subsection \ref{Se34},
the generalized Wigner operator is introduced and its properties are discussed.
In Sections \ref{Se4} and \ref{Se5}, we find an explicit form of the generalized Wigner
operators in the case of massless representations with continuous (infinite) spin and
in the case of helicity representations, respectively. Then with the help
of these operators we construct the corresponding local relativistic fields.
We also demonstrate here that in the case of helicity representations, the approach
under consideration automatically leads to a description of the corresponding
relativistic fields in terms of potentials defined up to gauge transformations.

\setcounter{equation}0
\section{Little group and the fundamental Wigner operators}\lb{Se2}
\setcounter{equation}0

\quad \  In this section we briefly describe the basic notation and conventions used in
 the paper.

We begin with the well known one-to-one correspondence between the set of coordinates $x_{\mu}\in\mathbb{R}^{1,3}$
and the set $\mathcal{H}$ of Hermitian matrices $X=x_{\mu} \sigma^{\mu} \in \mathcal{H}$,
 where $\sigma^0 =I_2$ is the $2 \times 2$ unit matrix and $\sigma^i$, $i=1,2,3$ are the Pauli $\sigma$-matrices. The action of the
group $\mathrm{SL}(2,\mathbb{C})$ on the set $\mathcal{H}$
\be
\lb{sl-h-p}
X \to X' = A X A^{\dagger}\, , \qquad X, X' \in \mathcal{H}\, , \quad
A \in \mathrm{SL}(2,\mathbb{C})
\ee
leads to the following group homomorphism
 $\mathrm{SL}(2,\mathbb{C}) \to \mathrm{SO}^\uparrow(1,3)$:
\be
\lb{Alambd-p}
A\, \sigma^\mu A^{\dagger} = \sigma^\nu \; \Lambda_\nu^{\;\; \mu}(A)\,, \qquad A
\in \mathrm{SL}(2,\mathbb{C})\, , \quad \Lambda_\mu{}^{\nu}(A) \in \mathrm{SO}^{\uparrow}(1,3)
\,.
\ee

Let $p  \, \in \,{\mathbb{R}^{1,3}}$ be
the 4-momentum of the relativistic particle and let
$\overset{_{\mathrm{\;o}}}{p}  \, \in \,{\mathbb{R}^{1,3}}$ be some singled out test 4-momentum. The momentum coordinates $p_\mu$
and $\overset{_{\mathrm{\;o}}}{p}_\mu$ are transformed one into another by the Wigner operator $A_{(p)}$ according to the relation
\be \lb{fr4-p}
A_{(p)} (\overset{_{\mathrm{\;o}}}{p}\,  \sigma) A_{(p)}^{\dagger} = (p\,\sigma)  \quad \Rightarrow
\quad   \Lambda_\mu{}^{\nu}(A_{(p)}) \, \overset{_{\mathrm{\;o}}}{p}_{\nu} = p_{\mu} \, ,
\ee
where we have used the notation $(p \sigma) \,{: =}\, p_{\mu} \sigma^{\mu}$.
The transformations (\ref{fr4-p}) are in fact the definition of the operators $A_{(p)} \in \mathrm{SL}(2, \mathbb{C})$
which transform some test momentum $\overset{_{\mathrm{\;o}}}{p}$ into the arbitrary momentum $p$.
It is well known that the above transformations preserve the square of the 4-momentum $p_\mu p^\mu$, i.e. $p_\mu p^\mu=\overset{_{\mathrm{\;o}}}{p}_\mu \overset{_{\mathrm{\;o}}}{p}^\mu$.
Further we will consider the case of massive particles,
where the 4-momentum is time-like,
$p_\mu p^\mu=\bm{m}^2\neq 0$,
and the case of massless particles, when the 4-momentum is light-like,
$p_\mu p^\mu=0$.

Among the transformations of the Lorentz group  $\mathrm{SL}(2, \mathbb{C})$ there
are those that preserve the test momentum  $\overset{_{\mathrm{\;o}}}{p}$, i.e.:
\be \lb{fr5}
h (\overset{_{\mathrm{\;o}}}{p}\, \sigma) h^{\dagger} = (\overset{_{\mathrm{\;o}}}{p}\, \sigma) \,.
\ee
Such matrices $h \in \mathrm{SL}(2,\mathbb{C})$ form
the Wigner little group $G_{0}$ of the  test momentum $\overset{_{\mathrm{\;o}}}{p}  \, \in \,{\mathbb{R}^{1,3}}$.

As follows from (\ref{fr4-p}) and (\ref{fr5}), the Wigner operators $A_{(p)}$ are defined
up to right multiplication by the elements of $G_{0}$ and parametrize the coset space $\mathrm{SL}(2,\mathbb{C})/G_{0}$.
According to (\ref{Alambd-p})--(\ref{fr4-p}), the left action of the group $\mathrm{SL}(2, \mathbb{C})$ on this
coset is defined by
\be \lb{twostars}
A \, A_{(p)} = A_{(\Lambda p)} \, h_{A,p} \,,
\qquad h_{A,p} \in G_{0} \,.
\ee
This relation, rewritten as
\be
\lb{acgf1-p}
h_{A,p} = A_{(\Lambda p)}^{-1} \, A \, A_{(p)} \quad \Rightarrow
\quad h_{A,\Lambda^{-1} p} = A_{(p)}^{-1} \, A \,
A_{(\Lambda^{-1} p)} \, ,
\ee
determines the element $h_{A,p}$ of the Wigner little group $G_{0}$,
where the labels $A,p$ in the notation $h_{A,p}$ indicate its
dependence on $A \in \mathrm{SL}(2,\mathbb{C})$ and the 4-momentum $p$.

Unitary irreducible representations of the Poincar\'{e} group are induced by
unitary irreducible representations of the Wigner little group $G_{0}$,
the structure of which is determined by the mass parameter of the particle being considered,
i.e., massive particle with $\bm{m} \neq 0$ or massless one with $\bm{m}=0$.

$\bullet$ Massive case. The test momentum $\overset{_{\mathrm{\;o}}}{p}$ can be chosen as
the rest particle momentum
\be \lb{fr3-pm}
||\overset{_{\mathrm{\;o}}}{p}_{\mu}|| = (\overset{_{\mathrm{\;o}}}{p}_{0}, \overset{_{\mathrm{\;o}}}{p}_{1},
\overset{_{\mathrm{\;o}}}{p}_{2},
\overset{_{\mathrm{\;o}}}{p}_{3})= (\bm{m},0,0,0)
\ee
and the Wigner little group is a unitary group: $G_{0} \simeq \mathrm{SU}(2)$.
 In this case, formula (\ref{acgf1-p}) can also be written
 in the form
\be \lb{onestar}
h_{A,p} = (h_{A,p})^{-1 \dagger} = A^{\dagger}_{(\Lambda p)} A^{-1 \dagger} A^{-1 \dagger}_{(p)} \,.
\ee

$\bullet$  Massless case. The test momentum $\overset{_{\mathrm{\;o}}}{p} \, \in \,{\mathbb{R}^{1,3}}$ can be fixed as
the light-cone particle momentum
\be \lb{fr3-p}
||\overset{_{\mathrm{\;o}}}{p}_{\mu}|| = (\overset{_{\mathrm{\;o}}}{p}_{0}, \overset{_{\mathrm{\;o}}}{p}_{1},
\overset{_{\mathrm{\;o}}}{p}_{2},
\overset{_{\mathrm{\;o}}}{p}_{3})= (E,0,0,E) \, , \;\;\;\;
 \;\;\;\; {(\overset{_{\mathrm{\;o}}}{p}\, \sigma) =
 \begin{pmatrix}
 2E & 0\\
 0  & 0
 \end{pmatrix} \, .}
\ee
The Wigner little group is formed by the matrices
\be
\lb{hhhh}
h =  \begin{pmatrix} 1 & b_1 + {\sf i} b_2 \\ 0 & 1 \end{pmatrix} \,
\begin{pmatrix} e^{{\sf i}\theta/2} & 0  \\ 0 & e^{-{\sf i}\theta/2}
\end{pmatrix}\, ,
\ee
where $\theta \in [0,2\pi)$ and $\vec{b} =(b_1,b_2)
\in \mathbb{R}^2$, and is isomorphic to the group of motions of two-dimensional Euclidean space
$G_{0} \simeq \mathrm{ISO}(2)$.

\setcounter{equation}0
\section{Generalized
Wigner operators and relativistic fields}\lb{Se3}

\quad\  Irreducible representations of the Poincar\'{e} group as well as its covering  group
are classified according to the eigenvalues of the corresponding Casimir operators.
The Poincar\'{e} algebra in four dimensions has two Casimir operators:
the squared momentum $\hat{P}^2$ and the squared Pauli-Lubanski vector $\hat{W}^2$.
For the cases under consideration, all possible unitary irreducible representations are determined by the following eigenvalues of these operators \cite{Wig39,Wig48,BarWig48}.
\begin{itemize}
\item Massive representations. \\ On the spaces of these representations, the following conditions are fulfilled:
\be
\lb{mn0}
\hat{P}^2 = \bm{m}^2 \,, \qquad \hat{W}^2 = - \bm{m}^2 j(j+1) \,,
\ee
where $\bm{m}>0$ is the real parameter describing the particle mass, and $j \in \mathbb{Z}_{\geq0}/2$ is the particle spin.
\item Massless representations. \\ In these representations, the Casimir operators are equal to:
\be
\lb{mless}
\hat{P}^2 = 0\,, \qquad \hat{W}^2 = - \bm{\mu}^2 \,,
\ee
where $\bm{\mu} \in \mathbb{R}_{\geq0}$ is the mass dimension parameter.
Depending on the value of this parameter, the massless representations belong to one of the two classes, i.e. they are
subdivided into
\begin{itemize}
\item
either the helicity representations with $\bm{\mu}=0$;
\item
or the continuous (infinite) spin representations with $\bm{\mu} > 0$.
\end{itemize}
\end{itemize}
Note also that the parameter $\bm{\mu}$, which defines the massless continuous spin representations  (\ref{mless}), 
corresponds to the radius of the circle
in the auxiliary two-dimensional Euclidean space that arises when considering the small group  (\ref{hhhh}).
In addition, we emphasize that the continuous spin representation contains an infinite number
of massless states with all possible integer or half-integer
helicities\footnote{For these representations helicities are not
invariants under the action of the Lorentz group $\mathrm{SL}(2,\mathbb{C})$.}.
The name ``continuous spin representation'' was developed historically after its introduction in  \cite{Wig39,Wig48,BarWig48}
and is associated with the possibility of describing this discrete set as a Fourier series with respect to the angular variable
which is similar to the angle $\theta$ in (\ref{hhhh}).
To avoid any misunderstandings, we also use the name ``infinite spin representations'' for such representations.

\subsection{Bargmann-\!Wigner wave functions}\lb{Se31}

\quad\  The main method in constructing unitary irreducible representations $U$ 
of the Poincar\'{e} covering group $\mathrm{ISL}(2,\mathbb{C})$ is
based on the Wigner scheme \cite{Wig39,Wig48,BarWig48}. Within this scheme, 
all unitary irreducible representations (UIR's)
are realized on the  Bargmann-\!Wigner wave functions $\Phi_{M}(p)$, which
depend on the four-momentum $p_{\mu}$ on the mass shell $p^{\mu}p_{\mu} = \bm{m}^2$ and
have an additional (multi)index $M$  that runs over a finite or infinite number of values.
This index is related to the unitary irreducible representation of the Wigner little group $G_{0}$.

Representation $U(A)$ of the element
$A \in SL(2,\mathbb{C})$ acts on the Bargnmann-Wigner wave function as follows:
\be \lb{act-in-g}
\Phi^\prime_{M}(p) := [U(A) \Phi]_{M}(p) = \mathcal{D}_{MN}(h_{A,\Lambda^{-1}p})
\Phi_{N}(\Lambda^{-1} p) \, ,
\ee
where the matrix $\Lambda$ is related to $A$ by (\ref{Alambd-p}), $\mathcal{D}_{MN}$ is UIR of the Wigner little group $G_{0}$ and the element
$h_{A,\Lambda^{-1}p}$ is determined by relation (\ref{acgf1-p}). One can check directly that
$U(A\, B) = U(A)\, U(B)$. The representation (\ref{act-in-g}) of the Poincar\'{e} group $\mathrm{ISL}(2,\mathbb{C})$, constructed in this way,
is induced by the unitary representation of the Wigner little group $G_{0}$.

However, the transformation law (\ref{act-in-g}) becomes non-local for the space-time field (field in coordinate representation)
obtained from the Bargmann-\!Wigner wave function $\Phi_{M}(p)$ using the Fourier transform.
To find local relativistic fields, it is necessary to modify the Bargmann-\!Wigner wave function before performing the Fourier transform.
In the next subsection we will briefly describe how it can be done for the case of massive representations
(for a detailed discussion of them, see  \cite{StW,IsPod0,IsPod1}).

\subsection{Massive representations\label{masrep}}

\quad \ In the massive case, UIR of the Wigner little group $G_{0} = \mathrm{SU}(2)$ is characterized by
the spin quantum number $j \in \mathbb{Z}_{\geq0}/2$.
This representation is realized on the Bargmann-Wigner wave functions $\Phi_{a_1 \cdots a_{2j}}(p)=\Phi_{(a_1 \cdots a_{2j})}(p)$ which
are the $\mathrm{SU}(2)$-spinors of rank $2j$, where $a =1,2$ is the $\mathrm{SU}(2)$-index and the
parentheses $(a_1 \dots a_{2j})$ mean total symmetrization over all indices.
In this case, the transformation law (\ref{act-in-g}) is written as
\be \lb{act-in-gm}
\Phi^\prime_{a_1 \dots a_{2j}}(p):=[U(A)\Phi]_{a_1 \dots a_{2j}}(p) = (h_{A,\Lambda^{-1}p})_{a_1}^{\; b_1} \cdots  (h_{A,\Lambda^{-1}p})_{a_{2j}}^{\; b_{2j}}\; \Phi_{b_1 \dots b_{2j}}(\Lambda^{-1}p)\, .
\ee

Let us move from the Bargmann-Wigner wave functions $\Phi_{a_1 \cdots a_{2j}}(p)$
to the new wave functions whose transformation law, in contrast to the transformation law (\ref{act-in-gm}), uses the matrices which do not depend
on the argument $p$. These new wave functions are the multicomponent fields labeled by the indices $q,\,r$ and defined as follows:
\be \lb{n-wf-f}
\Psi^{(q,r)}{}^{{\phantom{(}}\!\! \dot{\beta_1}...\dot{\beta_r}}_{{\phantom{(}}\!\!\alpha_1...\alpha_q}(p) =
(A_{(p)})_{\alpha_1}^{\, b_1} \cdots  (A_{(p)})_{\alpha_{q}}^{\, b_{q}} \,
(\tilde{A}^{-1 \dagger}_{(p)})^{\dot{\beta}_1 b_{q+1}} \cdots   (\tilde{A}^{-1 \dagger}_{(p)})^{\dot{\beta}_r b_{2j}}
\Phi_{b_1 \dots b_{2j}}(p) \,,
\ee
where $q=0,1,\ldots,2j$, $r=2j-q$ and
the following notation\footnote{In addition to the relativistic Pauli matrices $(\sigma^\mu)_{\alpha \dot{\beta}}$ we
also use the matrices $(\tilde{\sigma}^\mu)^{\dot{\alpha} \beta}=(\tilde{\sigma}^0,\tilde{\sigma}^{i}) =(\sigma^0,-\sigma^i)$.}
$\tilde{A}_{(p)}^{-1 \dagger}:= A_{(p)}^{-1 \dagger} \tilde{\sigma}_0$ is used.
The constructed fields (\ref{n-wf-f}) are symmetric with respect to indices of the same type:
$\Psi^{(q,r)}{}^{{\phantom{(}}\!\! \dot{\beta_1}...\dot{\beta_r}}_{{\phantom{(}}\!\!\alpha_1...\alpha_q}(p) =
\Psi^{(q,r)}{}^{ (\dot{\beta_1}...\dot{\beta_r})}_{(\alpha_1...\alpha_q)}(p)$.
An important property of the fields (\ref{n-wf-f}) is their transformation law
\be \lb{act-in-gmg}
[U(A) \Psi^{(q,r)}]{}^{{\phantom{(}}\!\! \dot{\beta_1}...\dot{\beta_r}}_{{\phantom{(}}\!\!\alpha_1...\alpha_q}(p)
= A^{\;\; \gamma_1}_{\alpha_1} \cdots A^{\;\; \gamma_q}_{\alpha_q}\;\;
\bigl(A^{\dagger -1} \bigr)^{\dot{\beta_1}}_{\;\; \dot{\kappa}_1} \cdots
\bigl(A^{\dagger -1} \bigr)^{\dot{\beta_r}}_{\;\; \dot{\kappa}_r}
\;\; \Psi^{(q,r)}{}^{{\phantom{(}}\!\! \dot{\kappa}_1...\dot{\kappa}_r}_{{\phantom{(}}\!\! \gamma_1...\gamma_q}(\Lambda^{-1} \cdot p)  \; ,
\ee
which is a direct consequence of the transformation law (\ref{act-in-gm}) and relations
(\ref{acgf1-p}), (\ref{onestar}). Here $A \in \mathrm{SL}(2,\mathbb{C})$.
Since the transformation matrix in (\ref{act-in-gmg}) does not depend on the 4-momentum $p$,
the constructed fields
$\Psi^{(q,r)}{}^{{\phantom{(}}\!\! \dot{\beta_1}...\dot{\beta_r}}_{{\phantom{(}}\!\!\alpha_1...\alpha_q}(p)$ are the standard local spin-tensor fields
in the momentum representation. The operators  $A_{(p)}^{\otimes q} (\tilde{A}_{(p)}^{-1 \dagger})^{\otimes r}$, $r+q=2j$,
which transform the Bargmann-Wigner wave functions $\Phi_{a_1 \cdots a_{2j}}(p)$ into local relativistic fields
$\Psi^{(q,r)}{}^{{\phantom{(}}\!\! \dot{\beta_1}...\dot{\beta_r}}_{{\phantom{(}}\!\!\alpha_1...\alpha_q}(p)$,
{are called the spin $j$ Wigner operators}.

The fields $\Psi^{(q,r)}{}^{{\phantom{(}}\!\! \dot{\beta_1}...\dot{\beta_r}}_{{\phantom{(}}\!\!\alpha_1...\alpha_q}(p)$
at fixed spin $j=(q+r)/2$ are not independent of mass-shell:
they automatically satisfy the Dirac-Pauli-Fierz equations
\begin{equation}\label{osudp}
\begin{array}{l}
(p\,\tilde{\sigma})^{\dot{\gamma}_1 \alpha_1}
\Psi^{(q,r)}{}^{{\phantom{(}}\!\! \dot{\beta_1}...\dot{\beta_r}}_{{\phantom{(}}\!\!\alpha_1...\alpha_q}(p)=
\bm{m} \, \Psi^{(q-1,r+1)}{}^{{\phantom{(}}\!\! \dot{\gamma}_1 \dot{\beta_1}...\dot{\beta_r}}_{{\phantom{(}}\!\! \alpha_2...\alpha_q}(p) \; ,
\;\;\;\; (r=0,\dots,2j-1) \; , \\[0.2cm]
(p\,\sigma)_{\gamma_1 \dot{\beta}_1}
\Psi^{(q,r)}{}^{{\phantom{(}}\!\! \dot{\beta_1}...\dot{\beta_r}}_{{\phantom{(}}\!\!\alpha_1...\alpha_q}(p)
= \bm{m} \, \Psi^{(q+1,r-1)}{}^{{\phantom{(}}\!\! \dot{\beta_2}...\dot{\beta_r}}_{{\phantom{(}}\!\! \gamma_1 \alpha_1...\alpha_q}(p) \; ,
\;\;\;\; (r=1,\dots,2j) \; .
\end{array}
\end{equation}
{which follow from (\ref{fr4-p}) and
(\ref{fr3-pm}). In particular, for $j=\frac{1}{2}$,
eqs. (\ref{osudp}) are two Weyl projections of the
spin-$\frac{1}{2}$ Dirac equation.}
These and other aspects of such a description of massive particle states are given in detail in \cite{IsPod0,IsPod1}.
{Other methods of deducing 
relativistic field equations
from the first principles were also considered in \cite{BekBoul}.}

\subsection{Massless representations}\lb{Se33}

\quad \ UIR's of the group $\mathrm{ISO}(2)$, being the Wigner little group $G_{0}$
in the massless case, are well known (see e.g. \cite{Vilen,ZhSh}).
They are characterized by a real nonnegative parameter $\bm{\rho}$ and realized on the
functions $\Phi(\varphi)$ defined on the circle of the radius $\bm{\rho}$
on the plane parameterized by the vector coordinates $\vec{t}_{\varphi}=((t_{\varphi})_1,(t_{\varphi})_2)=
(\bm{\rho}\cos\varphi,\bm{\rho}\sin\varphi)$.
Induced UIR's of the Poincar\'{e} group $\mathrm{ISL}(2,\mathbb{C})$ are realized on the {wave} functions $\Phi(p,\varphi)$,
which also depend on the 4-momentum $p_\mu$.
In the massless case, the formula (\ref{act-in-g})
of $\mathrm{SL}(2,\mathbb{C})$ takes the following form:
\be \lb{act-in-gml}
\begin{array}{rcl}
\Phi^\prime(p,\varphi)&:=&\displaystyle
[U(A) \Phi](p,\varphi)  = \int\limits_{0}^{2\pi} d \varphi^\prime \  \mathcal{D}_{\varphi\varphi^{\prime}}(
\theta_{A,\Lambda^{-1} p},
\vec{b}_{A,\Lambda^{-1}p})
\, \Phi (\Lambda^{-1} p , \varphi^{\prime}) \\[6pt]
&\equiv&\displaystyle
 e^{  - i \, \vec{b}_{A,\Lambda^{-1}p} \cdot  \vec{t}_{\varphi}} \, \Phi (\Lambda^{-1} p , \varphi-\theta_{A,\Lambda^{-1}p}) \,,
\end{array}
\ee
where the operator $\mathcal{D}_{\varphi\varphi^{\prime}}(\theta_{A,\Lambda^{-1} p}, \vec{b}_{A,\Lambda^{-1} p})$
is the representative of the little group element $h_{A,\Lambda^{-1} p} \in \mathrm{ISO}(2)$ in unitary representation (see e.g. \cite{Vilen}).
The dependence of the element $h \in \mathrm{ISO}(2)$ (paramet\-rized by the pair $(\theta, \vec{b})$ as in  (\ref{hhhh}))
on $A$, $p$ was discussed in \cite{BIPF}.
The matrix element $\mathcal{D}_{\varphi \varphi^\prime}(\theta,\vec{b})$ has the following explicit form:
\be \lb{D-a-p}
\mathcal{D}_{\varphi \varphi^{\prime}}(\theta, \vec{b}) =
e^{  - i \, \vec{b}_{\mathstrut} \cdot  \vec{t}_{\varphi}}\delta(\varphi-\varphi^{\,\prime}-\theta) \, ,
\ee
where $\delta(\varphi)$ is the periodic Dirac $\delta $-function on a circle and
$
\vec{b}_{\mathstrut} \cdot  \vec{t}_{\varphi} = \bm{\rho} \, \bigl( b_1 \cos \varphi+b_2 \sin \varphi \bigr)
$.

There is another equivalent formulation of the induced massless Poincar\'{e} group representations (see e.g. \cite{BekSk}, \cite{BIPF}),
which is obtained after expanding the Bargmann-\!Wigner wave function $\Phi(p,\varphi)$ into a Fourier series:
\be \lb{Phi-exp-p}
\Phi(p,\varphi)
= \sum_{n=-\infty}^{\infty} \Phi_{n}(p)
 e^{i n \varphi} \,.
\ee
In the space of functions $\Phi_{n}(p)$, the transformation law (\ref{act-in-gml}) has the form:
\be \lb{rwd2-p}
[U(A) \Phi]_{n}(p) =  \sum_{m=-\infty}^{\infty}  \mathcal{D}_{nm} (\theta_{A,\Lambda^{-1} p},
\vec{b}_{A,\Lambda^{-1} p}) \, \Phi_m (\Lambda^{-1} p) \, ,
\ee
where the matrix elements $\mathcal{D}_{nm}(\theta, \vec{b})$
of the little group $\mathrm{ISO}(2)$ are
\be \lb{mn-b-p}
 \mathcal{D}_{nm} ( \theta, \vec{b}) =
 (- i e^{i \beta})^{m-n} \, e^{- i m \theta} J_{(m-n)}(b{\bm{\rho}}) \; .
\ee
Here the real numbers $\beta$ and $b$ are the polar coordinates of the 2-vector
$\vec{b} =(b_1,b_2)= b \, (\cos \beta, \sin \beta)$ and $J_{(n)}(x)$ are the Bessel functions of integer order $n$.

The induced unitary Wigner
representations (\ref{act-in-gml}) and (\ref{rwd2-p}) of the Poincar\'{e} group
$\mathrm{ISL}(2,\mathbb{C})$  are irreducible for
$\bm{\rho} \neq 0$ and infinite-dimen\-sional.
The corresponding representations ${\cal D}$ of $\mathrm{ISO}(2)$
written for the discrete bases (\ref{Phi-exp-p})-(\ref{mn-b-p})
are decomposed into a direct sum of one-dimensional
irreducible representations of the subgroup
$\mathrm{SO}(2) \subset \mathrm{ISO}(2)$.
The action of the group $\mathrm{ISO}(2)$ in the
representation ${\cal D}$ in the limit $\bm{\rho} \to 0$ is
reduced to the action of its subgroup $\mathrm{SO}(2)$.
Thus, in the limit $\bm{\rho} \to 0$, the representation
${\cal D}$  of $\mathrm{ISO}(2)$ becomes reducible and
is decomposed into a sum of infinite number of one
dimensional UIR's.
In the limit $\bm{\rho} \to 0$ expression (\ref{rwd2-p})
 is used in the description
of the $\mathrm{ISL}(2,\mathbb{C})$ helicity representations with finite numbers of relativistic spin states (see Section
{\bf \ref{Se5}} below).

To construct a local relativistic field corresponding to the Bargmann-\!Wigner wave function $\Phi(p, \varphi)$, we use a generalization
of the method applied in Subsection {\bf \ref{masrep}} in the case of massive representations.

\subsection{Generalized Wigner operators in the infinite spin case} \lb{Se34}

\quad \ In the massive case,
relation (\ref{n-wf-f}) between the local relativistic fields and the Bargmann-\!Wigner wave functions is
{served} by the Wigner operators
$A_{(p)}^{\otimes q} (\tilde{A}_{(p)}^{-1 \dagger})^{\otimes r}$.
In the general case of massless representations including infinite-dimensional representations of continuous spin,
it is required to use an infinite-dimensional generalization of this correspondence.

The main role in this correspondence is played by the Wigner operator $A_{(p)}$ whose matrices parameterize the coset-space $\mathrm{SL}(2,\mathbb{C})/G_{0}$.
After restoring the indices, {the left action (\ref{twostars}) of the element
$A \in \mathrm{SL}(2,\mathbb{C})$ on $\mathrm{SL}(2,\mathbb{C})/G_{0}$ is written as}
\be \lb{act-sl}
A_{\alpha}{}^{\beta} \, (A_{(p)})_{\beta}{}^{c} = (A_{(\Lambda p)})_{\alpha}{}^{b} \,( h_{A,p})_{b}{}^{c}\,,
\ee
where the matrix {$(A_{(p)})_{\alpha}{}^{a}$
of the Wigner operator  has the $ \mathrm{SL}(2,\mathbb{C})$-index $\alpha$
and $G_{0}$-index $a$ (in particular, for the massive case, we have $G_{0}=\mathrm{SU}(2)$).}
Thus, the operators $A_{(p)}^{\otimes q} (\tilde{A}_{(p)}^{-1 \dagger})^{\otimes r}$ play the role of a bridge that
converts the {little sub}group indices $a,b,...$
of  $G_{0}$-type into the relativistic
indices $\alpha,\beta,...$ of $\mathrm{SL}(2,\mathbb{C})$-type.

An analogue of such a transformation for the Bargmann-\!Wigner wave function $\Phi(p,\varphi)$ introduced in (\ref{act-in-gml})
{for the massless case}
is given by the following formula:
\be \lb{unpgr4}
\Psi(p, y) \ = \ \int\limits_{0}^{2\pi} d \varphi \, \mathcal{A}(p,y,\varphi)\, \Phi (p, \varphi) \, ,
\ee
where additional variables $y$ {denote a certain set of variables that play the role
of the relativistic $\mathrm{SL}(2,\mathbb{C})$-type indices while $\varphi$ is of the $\mathrm{ISO}(2)$-type
($G_{0}$-type) variable}.
The transform (\ref{unpgr4}) can be thought as an infinite-dimensional version of the transition (\ref{n-wf-f}):
it converts the index $\varphi$ of the Wigner little group $\mathrm{ISO}(2)$ into some (as yet undefined) relativistic index $y$.
{We call the integral operator with the kernel
$\mathcal{A}(p,y,\varphi)$ in the right-hand side of
(\ref{unpgr4})  {\it the generalized Wigner operator}.}

In the next section, {the $\mathrm{SL}(2,\mathbb{C})$-type variables $y$ are taken to be the
components of the Lorentz vector $\eta^\mu \in \mathbb{R}^{1,3}$,
or the components of a  pair of Weyl spinors $u^{\alpha}$, $\bar{u}^{\dot\alpha}$.}
The requirement {of the locality
of the $\mathrm{SL}(2,\mathbb{C})$ transformations of the relativistic fields $\Psi(p,\eta)$, or $\Psi(p,u,\bar{u})$,}
leads to the definition of the explicit form of the generalized Wigner operators in (\ref{unpgr4}) for these two cases.

\setcounter{equation}0
\section{Relativistic fields with infinite spin} \lb{Se4}
\subsection{Additional vector
variables\label{advec}}

\quad \ {Let us consider variables $y$ in  (\ref{unpgr4}) as additional vector coordinates $\eta_\mu \in \mathbb{R}^{1,3}$ of the fields.
The field $\Psi(p,\eta)$ must have the local transformation law under the action of $A \in \mathrm{SL}(2,\mathbb{C})$ (independent of $p$), as
it was in (\ref{act-in-gmg})}.
In the case when, a relativistic field is characterized
by its dependence on the auxiliary vector variables
$\eta_\mu$ instead of external indices, this transformation law takes the form
\be \lb{unpgr41}
\Psi^\prime(p, \eta) \ := \ [U(A) \Psi](p, \eta) \ = \ \Psi(\Lambda^{-1} p , \Lambda^{-1} \eta) \, ,
\ee
where the matrices $A$ and $\Lambda$ are related by (\ref{Alambd-p}) and momentum $p$ is on the
mass shell $p^2=0$.
Taking into account (\ref{act-in-gml}) and (\ref{unpgr4}),
the condition (\ref{unpgr41}) leads to the equation for the kernel
$\mathcal{A}(p,\eta,\varphi)$:
\be \lb{unpgr7aa}
\mathcal{A}(\Lambda^{-1} p, \Lambda^{-1} \eta, \varphi) \ = \ \int d\varphi^\prime
\mathcal{A}(p,\eta,\varphi^\prime)\,
\mathcal{D}_{\varphi^\prime\varphi}(\theta_{A,\Lambda^{-1} p},\vec{b}_{A,\Lambda^{-1} p}) \, .
\ee
This equation {on the generalized Wigner operator}
is an infinite-dimensional counterpart of the matrix coset transformation (\ref{act-sl}).

As shown in  \cite{ShTor1,ShTor2,BIPF},
the infinitesimal form of a relation (\ref{unpgr7aa}) is represented as three
differential equations for the kernel $\mathcal{A}(p,\eta,\varphi)$,
which can be solved exactly.
There are two types of solutions of these  equations, which  can be called non-singular and singular \cite{ShTor1,ShTor2,BIPF}.

\begin{itemize}
\item Non-singular solution. \\
The operator $\mathcal{A}(p,\eta,\varphi)$ that satisfies (\ref{unpgr7aa}) has the following explicit form:
\be \lb{sol3}
\mathcal{A}(p,\eta,\varphi) = e^{\,i {\bm{\mu}} \,  \eta\cdot \varepsilon_{(1)}(\varphi) /(\eta \cdot p)} \, f(\eta\!\cdot\! \eta,  \eta \! \cdot\! p) \, ,
\ee
where $f(\eta\! \cdot\! \eta,  \eta\! \cdot\! p)$ is an arbitrary function  of two variables
{$\eta\! \cdot\! \eta := \eta^\mu \eta_\mu$ and
$\eta\! \cdot\! p:= \eta^\mu p_\mu$}.
The dimensional constant $\bm{\mu}$ is related to the dimensionless constant $\bm{\rho}$ by the relation
\be \lb{mu-rho}
\bm{\mu} =E\bm{\rho} \, ,
\ee
where $E$ is defined by nonzero components of the massless test momentum $\overset{_{\mathrm{\;o}}}{p}$ in (\ref{fr3-p}).
In (\ref{sol3}), the vector $\varepsilon^\mu_{(1)}(\varphi)=\varepsilon^\mu_{(1)}\cos\varphi- \varepsilon^\mu_{(2)}\sin\varphi$
is constructed from the polarization vectors $\varepsilon_{(1)}$ and $\varepsilon_{(2)}$
which are determined by the conditions
$\varepsilon_{(1)}\!\cdot\!\varepsilon_{(1)}  =  \varepsilon_{(2)}\!\cdot\!\varepsilon_{(2)} = -1$,
$p\!\cdot\!\varepsilon_{(1)} = p\!\cdot\!\varepsilon_{(2)}  = \varepsilon_{(1)}\!\cdot\!\varepsilon_{(2)} = 0$.
\end{itemize}

\begin{itemize}
\item Singular solution. \\ 
In this case, the solution of (\ref{unpgr7aa}) can be written as follows:
\be \lb{sol-2d}
\mathcal{A}(p,\eta,\varphi) \ = \
\delta(\eta\!\cdot\!p) \, \delta(\eta\!\cdot\!{\varepsilon}_{(2)}(\varphi))\,
e^{i {\bm{\mu}} \, \eta \cdot {\varepsilon}  /
(\eta\cdot {\varepsilon}_{(1)}(\varphi) )}  \,
g(\eta\!\cdot\!{\varepsilon}_{(1)}(\varphi)) \, ,
\ee
where $g(\eta\!\cdot\!{\varepsilon}_{(1)}(\varphi))$ is an arbitrary function and $\varepsilon^\mu_{(2)}(\varphi)=\varepsilon^\mu_{(1)}\sin\varphi+ \varepsilon^\mu_{(2)}\cos\varphi$.
In (\ref{sol-2d}), the vector ${\varepsilon}^\mu$  obeys the conditions
$\varepsilon\!\cdot\! p  = 1$, $\varepsilon\!\cdot\!\varepsilon =\varepsilon\!\cdot\!\varepsilon_{(1)}  = \varepsilon\!\cdot\!\varepsilon_{(1)} = 0$ and forms tetrads with the vectors $p$, $\varepsilon_{(1)}$, $\varepsilon_{(2)}$.

\end{itemize}

{Two expressions (\ref{sol3}) and (\ref{sol-2d}) for the generalized Wigner operators
produce two types of relativistic fields $\Psi(p,\eta)$, according to the transform (\ref{unpgr4}), where one can use
the expansion (\ref{Phi-exp-p}) for the Bargmann-\!Wigner wave function.}
By construction, the fields $\Psi(p, \eta)$ describe physical states with the eigenvalues (\ref{mless})
of the Casimir operators. {Indeed, the first
eigenvalue in (\ref{mless}) is evident. The second eigenvalue was checked in \cite{BIPF}, where we used
the explicit formula for the square
of the Pauli-Lubanski operator $\hat{W}^2$ acting
in the space of the fields $\Psi(p, \eta)$.}
All these statements are fulfilled for both non-singular and singular  solutions.

\subsection{Additional spinor variables}\lb{Se42}

\quad \ Let us now consider the case when the components of the commuting Weyl spinor
$u_\alpha$, $\bar{u}_{\dot{\alpha}}=(u_{\alpha})^ {\dagger} $,
$\alpha, \dot{\alpha} = 1, 2$ are taken as additional variables $y$ in (\ref{unpgr4}).
The local relativistic field $\Psi(p,u,\bar{u})$ constructed by using the generalized Wigner operator
in the procedure described above was found in \cite{BIPF} and has the following form:
\be \lb{dsp2-sp}
\Psi(\pi,\bar\pi,u,\bar{u})  =  \int\limits_{0}^{2\pi} d \varphi \, \exp \left\{-i  \, {\bm{\mu}} \,
\Bigl ( \frac{ u^{\alpha} \lambda_{\alpha}}{u^{\beta} \pi_{\beta}} \, e^{- i \varphi} +
\frac{ \bar{u}^{\dot{\alpha}} \bar{\lambda}_{\dot{\alpha}}}{\bar{u}^{\dot{\beta}} \bar{\pi}_{\dot{\beta}}} \, e^{ i \varphi} \Bigr ) \right\}
\mathrm{f}(u^{\gamma} \pi_{\gamma} \, e^{\frac{i}{2} \varphi}, \bar{u}^{\dot{\gamma}} \bar{\pi}_{\dot{\gamma}} \, e^{-\frac{i}{2} \varphi} )\, \Phi (p, \varphi)\,,
\ee
where $\mathrm{f}$ is an arbitrary function of two conjugated scalar variables.
In (\ref{dsp2-sp}), the commuting Weyl spinor
$\pi_{\alpha}$ is actually a twistor which defines massless 4-momentum through
the Cartan-Penrose relation
\be \lb{seo12}
(p \,\sigma)_{\alpha\dot\beta} =
\bigl( A_{(p)} \, (\overset{_{\mathrm{\;o}}}{p} \,\sigma)
\, A_{(p)}^\dagger \bigr)_{\alpha\dot\beta}
= \pi_{\alpha} \bar{\pi}_{\dot{\beta}} \, .
\ee
Another commuting Weyl spinor $\lambda_{\alpha}$ is determined
by the normalization condition
$\pi^{\alpha} \lambda_{\alpha} = 1$, that is, a pair of spinors
$\pi_{\alpha}$, $\lambda_{\alpha}$ forming the basis
in the 2-component spinor space. {Thus, for the choice (\ref{fr3-p}), one can take the parametrization:
$A_{(p)} =  \begin{pmatrix}
\pi_1/\sqrt{2E} \, , & \lambda_1 \sqrt{2E}\\
\pi_2/\sqrt{2E} \, , & \lambda_2 \sqrt{2E} \\
\end{pmatrix}$.}

The square of the Pauli-Lubanski vector $\hat{W}^2$ on the space of the fields
$\Psi(\pi,\bar{\pi},u,\bar{u})$ has the following
representation (see e.g. \cite{BuchIFKr}):
\be \lb{seo13a}
\hat{W}^2 = (u^{\alpha} \pi_{\alpha} \bar\pi_{\dot{\alpha}} \bar{u}^{\dot{\alpha}} ) \Bigl( \frac{\partial}{\partial \bar{u}^{\dot{\beta}}} \,\bar\pi^{\dot{\beta}} \pi^{\beta} \frac{\partial}{\partial u^{\beta}} \Bigr )\,.
\ee
By making use of the explicit form (\ref{dsp2-sp}) for $\Psi(\pi,\bar\pi,u,\bar{u}) $, it can be shown that
\be \lb{seo14-tw}
\hat{W}^2 \, \Psi(\pi,\bar\pi,u,\bar{u}) = -{\bm{\mu}}^2 \,  \Psi(\pi,\bar\pi,u,\bar{u}) \, .
\ee
Therefore, the UIR's of the group
$\mathrm{ISL}(2,\mathbb{C})$ with infinite spin are realized
on the field $\Psi(\pi,\bar\pi,u,\bar{u})$. For more details about the equations of motion for
the field $ \Psi(\pi,\bar\pi,u,\bar{u})$ see \cite{BFIR,BFI,BIPF}.

\setcounter{equation}0
\section{Relativistic gauge fields} \lb{Se5}

\quad \ Another  {important}  example of massless UIR's of the
Poincar\'{e} covering group is massless helicity
representations. Such representations are {physically motivated}
and wildly used in higher spin field theory (see, e.g., the review \cite{hsrev}).

Helicity representations are induced
from UIR's of the subgroup $\mathrm{SO}(2) \subset \mathrm{ISO}(2)$,
{or from UIR's of $\mathrm{ISO}(2)$
with trivial realizations
of the 2-dimensional translations.} All UIR's of the $\mathrm{SO}(2)$ group are one-dimensional
and characterized by an integer or half-integer number corresponding to particle helicity.
In what follows we will
be interested only in representations with integer helicities
and demonstrate that the approach of the generalized Wigner operators \cite{BIFPP}
reproduces well-known expressions for massless fields of higher helicities.
One of the advantages of this method is obtaining a group theoretical description of massless helicity
representations in terms of field potentials defined up to gauge transformations.

Helicity representations are induced by the  UIR's of the Wigner little group $ISO(2)$ at $\bm{\rho} = 0$.
In this limit, the transformation low (\ref{rwd2-p}) takes the form
\be \lb{rwd1-a}
[U(A) \Phi]_{n}(p) = e^{-  i n \theta_{\!\!A,\Lambda^{-1} p}} \, \Phi_n (\Lambda^{-1} p) \, , \;\;\;\;\;
 \forall \; n \in \mathbb{Z} \; ,
\ee
where the property $J_{m-n}(0) = \delta_{m,n}$  for the Bessel functions was used.
Thus, at $\bm{\rho} = 0$ the unitary representations of the Poincar\'{e} group are
decomposed into a direct sum {of UIR's induced from the
one-dimensional UIR's of $SO(2)$} which are described by the functions $\Phi_n(p)$.
{Here the number $n$ corresponds to particle helicity.}

Relativistic local fields are constructed from
the Wigner wave functions $\Phi_n(p)$ by using the transform (\ref{unpgr4})
{written in the discrete basis (\ref{Phi-exp-p}).
It turns out that} to construct a massless relativistic field $\Psi_n(p)$ with spin $n>0$,
two Bargmann-\!Wigner wave functions with opposite helicities $n$ and $-n$ should be used.
In fact, this corresponds to the construction of a massless irreducible representation of the extended
Poincar\'{e} group, which includes discrete transformations $P$ and $T$ (see e.g. \cite{GS,BGS}).
Thus, in accordance with  (\ref{unpgr4}), the local field of the helicity $n$ particle
has the form\footnote{The case $n=0$ is considered separately.}
\be
\lb{rfield}
\Psi_n(p,\eta) \ = \
\mathcal{A}(p,\eta,n) \Phi_n (p)  \ +  \ \mathcal{A}(p,\eta,-n)
\Phi_{-n} (p) \, , \quad n \in \mathbb{Z}_{>0} \,.
\ee
The explicit form of the generalized Wigner operator $\mathcal{A}(p,\eta,n)$ in the discrete basis for the case $\bm{\rho} = 0$
was found in \cite{BIFPP} and has the form
\be
\lb{wo.fs}
\mathcal{A}(p,\eta,n) = \left\{
\begin{array}{lll}
\displaystyle
\delta(\eta\! \cdot\! p) \, (\varepsilon_{(+)}\! \cdot\! \eta)^{n} \,   \;\;\;\;\; {\mbox{with}}\;\; n>0\, ,
\\ [10pt]
\displaystyle
\delta(\eta\! \cdot\! p) \, (\varepsilon_{(-)}\! \cdot\! \eta)^{-n} \,   \;\;\; {\mbox{with}}\;\; n<0\, ,
\end{array}
\right.
\ee
where $\varepsilon_{(\pm)}=  \varepsilon_{(2)} \pm i  \varepsilon_{(1)}$.
Then, {the substitution of (\ref{wo.fs}) into
(\ref{rfield}) gives the expression for the massless field of
the helicity $n$}:
\be
\lb{rwd7}
\Psi_n(p,\eta) \ = \ \delta(\eta\! \cdot\! p) \,F_n(p,\eta) \,
, \ee
where
\be \lb{rwd8a} F_n(p,\eta) \ = \ F^{_{(+)}}_{n} (p,
\eta) \ + \ F^{_{(-)}}_{n} (p, \eta) \, , \qquad F^{_{(\pm)}}_{n}
(p, \eta) = (\varepsilon_{(\pm)} \cdot \eta)^{n} \, \Phi_{\pm
n}(p)\, .
\ee

The explicit form of the functions (\ref{rwd8a}) reproduces
automatically the equations of motion of the field $F_{n} (p,
\eta)$:
\be
\lb{rwd9}
p^2 \, F_{n} (p, \eta)   =  0 \, , \quad
\left(p \cdot\! \frac{\partial}{\partial \eta}\right) \, F_{n} (p, \eta) = 0 \, , \quad
\left(\frac{\partial}{\partial \eta} \cdot\! \frac{\partial}{\partial \eta}\right) \, F_{n} (p, \eta) = 0 \, ,
\ee
\be
\lb{rwd12}
\left(\eta \cdot\! \frac{\partial}{\partial \eta}\right) \, F_{n} (p, \eta) = n \, F_{n} (p, \eta)  \, .
\ee
where the properties of polarization vectors were  used.
The last equation determines the degree of homogeneity for the field
$F_{n} (p, \eta)$ in the variables $\eta$. In addition, the
presence in the definition of $\Psi_{n} (p, \eta)$ of the field
$F_{n} (p, \eta)$ together with the $\delta$-function $\delta (\eta
\cdot p)$ leads to the following equivalence relation:
\be
\lb{rwd13}
F_{n} (p, \eta) \ \ \sim \ \ F_{n} (p, \eta)  + (p \cdot
\eta) \, \epsilon_{n-1} (p,\eta) \, ,
\ee
where 
{$\epsilon_{n-1} (p,\eta)$ are homogeneous functions in $\eta$ of degree $(n-1)$, such that they satisfy equations (\ref{rwd9}).}
Relation (\ref{rwd13}) is essentially a gauge transformation with the parameters $\epsilon_{n-1} (p,\eta)$.
Thus, the construction (\ref{rwd7})-(\ref{rwd8a}) reproduces the description of
gauge fields with all integer spins $n>0$ in terms of distributions (see \cite{BekSk,BeM}).
A method similar to that described in this section but using additional spinor
variables was discussed in \cite{ZFed}.

\setcounter{equation}0
\section{Conclusion}

\quad \

Let us briefly formulate the results.

In this paper, we have presented the main aspects of constructing a field realization of the $4D$ Poincar\'{e} group UIR's.
We considered all possible physically interesting cases of UIR's including massive and massless representations.
The basic fundamentals of our approach were the Wigner method for constructing induced representations.
At the same time, the key point was to find a relation between the {Bargmann-Wigner wave functions},
on which induced representations are implemented, and relativistic fields which are
locally transformed under the Poincar\'{e} group.

It is shown that such a connection is provided by the generalized Wigner operator, which explicitly relates
the Bargmann-Wigner wave functions and local relativistic fields.
In the massive case, this operator is described by the finite-dimensional matrix
and the local fields obtained are {automatically} satisfied by the well-known Dirac-Pauli-Fierz equations.
In the massless case including both helicity and infinite spin representations,
the generalized Wigner operator is either an infinite-dimensional
{matrix} or the kernel of the integral transform (\ref{unpgr4}).
Infinite spin fields are defined in the space with additional commuting coordinates which can be vectorial or spinorial.
In the case of vectorial additional coordinates,
we reproduce the infinite spin fields studied in \cite{ShTor1,ShTor2}.
In the case of spinor additional variables, we obtain twistor infinite spin fields \cite{BFIR,BFI}.
In the case of massless helicity representations,
the generalized Wigner operator contains the $\delta$-function that allows us to introduce local fields in form of potentials
which are defined up to gauge transformations.

\end{document}